\documentclass[a4paper,11pt]{article}

\usepackage{pos}
\usepackage{soul}

\title{$\bar{b}\bar{c}\,q_1q_2$ four-quark states from Lattice QCD}
\ShortTitle{$\bar{b}\bar{c}\,q_1q_2$ four-quarks}

\author[a]{M. Padmanath}
\author*[b]{Nilmani Mathur}

\affiliation[a]{Helmholtz Institut Mainz, Staudingerweg 18, 55128 Mainz, Germany
\\GSI Helmholtzzentrum fur Schwerionenforschung, Darmstadt (Germany)}

\affiliation[b]{Department of Theoretical Physics, Tata Institute of Fundamental Physics\\
  Homi Bhabha Road, Mumbai 400005, India}

\emailAdd{nilmani@theory.tifr.res.in}
\emailAdd{pmadanag@uni-mainz.de}

\abstract{We present the results of a lattice calculation of four-quark states with the quark contents $\bar{b}\bar{c}q_1q_2$ where $q_1,q_2  \in u,d,s$. For the spin 1 states, when the light quark ($q_1, q_2$) masses are lighter, we find at least one energy level below the possible elastic threshold energy levels. These calculations are performed on three dynamical $N_f = 2+1+1$ highly improved staggered quark ensembles at lattice spacings of about 0.12, 0.09, and 0.06 fm. We use the overlap action for light to charm quarks while a non-relativistic action with non-perturbatively improved coefficients is employed for the bottom quark.\\
  \flushright{TIFR-TH/21-19, MITP/21-051}
}

\FullConference{%
 The 38th International Symposium on Lattice Field Theory, LATTICE2021
  26th-30th July, 2021
  Zoom/Gather@Massachusetts Institute of Technology
}

\begin{document}
\maketitle

\section{Introduction}\label{Intro}
Hadron spectroscopy is passing through an incredible era with the discovery of a large number of subatomic particles in a short span of time \cite{Zyla:2020zbs}. Moreover, it is highly anticipated that many more hadrons will be discovered in the near future from the large data being collected and to be collected at LHCb, BESIII, BelleII, JPARC and JLAB. While some of the newly discovered particles could be understood as the regular mesons and baryons, many of them do not fit in the conventional valence quark-antiquark or three quark pictures. Observed decay channels for some of these states even manifestly demand interpretation through four or five valence quark configurations.
Though a number of theories have been put forward to interpret these new exotic states -- starting from compact tetraquarks, loosely bound molecules, hadroquarkonia and hybrids, etc., and even by non-resonant structures arising from threshold effects and triangle singularity, a coherent picture on the formations, structures and properties of these new states, is still elusive \cite{Ali:2017jda, Olsen:2017bmm, Guo:2017jvc, Karliner:2017qhf, Liu:2019zoy, Brambilla:2019esw}.

Lattice QCD is the best suitable way to study the energy spectra and structures of subatomic particles quantitatively with controlled systematic errors, which can further be improved with increasing computing resources. Naturally, lattice QCD is a promising tool to study these new particles. Here the primary goal is to understand the energy spectra and structures of these new particles and to predict many more such particles to guide in discovering them at experimental facilities. However, in order to unveil the infinite-volume physics from the finite-volume spectrum, it is 
essential to perform chiral and continuum extrapolations to the physical limits, along with a detailed 
finite-volume analysis to discern the pole distribution in the scattering amplitude across the complex energy plane. For the states such as $X(3872), Z(4430), P_c$-pentaquarks etc., where there are a number of multi-particle thresholds, it is 
quite challenging to obtain precise results with current-days available statistics and lattice technologies. As a result, only a few preliminary attempts have so far been made to investigate these newly discovered exotics {\it c.f.} in charmonia \cite{Prelovsek:2013cra,Padmanath:2015era,Prelovsek:2020eiw}. 

However, for the states well below the elastic threshold, lattice QCD can provide precise results to investigate them in detail and even can guide in discovering them. 
Indeed, a consistent prediction has been made by multiple lattice QCD
groups \cite{Bicudo:2016ooe, Francis:2016hui, Junnarkar:2018twb, PhysRevD.100.014503, PhysRevD.102.114506} where deeply bound four-quark states have been found with the
valence quark contents $\bar{b}\bar{b}ud$ and $\bar{b}\bar{b}us$
with spin $J = 1$ and isospin $I = 0$ and $I = 1/2$. An energy level of about 100-150 MeV
below the elastic threshold ($BB^*$) was found for $\bar{b}\bar{b}ud$
which can be associated with its binding energy assuming the finite volume effects are small in a system of two heavy mesons. For its strange-sibling, $\bar{b}\bar{b}us$, the respective threshold and the predicted binding energy
are $BB_s^*$ and 70-100 MeV. That makes these very states stable
under strong interactions, and hence they are the prime candidates to search
for at experimental facilities. However, a large center of momentum energy is required to produce a hadron with two bottom quarks and hence their discovery may get delayed. Other doubly heavy four-quarks that are also appealing are $\bar{b}\bar{c}ud$ $\bar{b}\bar{c}us$, $\bar{c}\bar{c}ud$ and $\bar{c}\bar{c}us$, and they have a greater discovery potential. Recently a state with the valence quark configuration $cc\bar u \bar d$, named as $T_{cc}^+$, has been discovered by LHCb \cite{LHCb:2021vvq, LHCb:2021auc}. Prior to that discovery, in a lattice calculation we found an energy level $23 \pm11 $ MeV below the elastic threshold, $DD^*$ \cite{Junnarkar:2018twb}. However, as mentioned earlier, it is necessary to make an amplitude analysis of finite-volume spectrum to associate that energy level with the discovered state.
For $\bar{c}\bar{c}us$, we also found an energy level $8\pm 8$ MeV below the elastic threshold $DD_s^*$, which again needs a detail volume study to find whether that is a scattering state or a bound state close to the threshold as in the case of $T_{cc}^+$.

Lattice calculations have also been performed recently for $\bar{b} \bar{c} u d  \, (J = 1, I = 0)$ and  $\bar{b} \bar{c} u s \, (J = 1, I = 1/2)$. In Ref. \cite{PhysRevD.99.054505} it was reported that for  $\bar{b} \bar{c} ud$, an energy level exists 15-61 MeV below its elastic threshold ($DB^*$). However, in Ref. \cite{PhysRevD.102.114506}  authors of Ref. \cite{PhysRevD.99.054505} refined their results with an improved quark source and reported a new set of results where no indication of any binding for either $\bar{b} \bar{c} u d$ or $\bar{b} \bar{c} u s$   was found. In Ref. \cite{Pflaumer:2021ong}, no binding for $\bar{b} \bar{c} u d/\bar{b} \bar{c} u s$ has also been reported.  $\bar{b} \bar{c} u d/\bar{b} \bar{c} u s$ have also been investigated through various models (see Ref. \cite{PhysRevD.99.054505} for references on such models). While the HQ-symmetry inspired and non-chiral models mostly found unbound or very weekly bound states, QCD sum rule and chiral models reported a bound state (both for 0 and 1-isospins) with binding energy over a wide range $\sim 20-400$ MeV.

Here we report preliminary results from our ongoing study of $T_{bc}(\bar{b} \bar{c} u d) (J=1, I= 0)$ and  $T_{bcs}(\bar{b} \bar{c} u s) (J=1, I= 1/2)$ channels. Contrary to the other two lattice calculations \cite{PhysRevD.102.114506, Pflaumer:2021ong} our preliminary results indicate the presence of an energy level, which can be clearly extracted,  below the respective elastic thresholds both for $\bar{b} \bar{c} u d$ and $\bar{b} \bar{c} u s$. We perform this calculation at both coarse and finer lattice spacings and find that the extracted energy levels have lattice spacing dependence -- the lowest energy level gets deeper below the elastic threshold at finer lattices.

\section{Lattice set up and operators}

The lattice set up for this work is the same as in ILGTI's previous calculations on heavy hadrons \cite{PhysRevD.99.031501, PhysRevLett.121.202002}, four-quarks \cite{Junnarkar:2018twb} and dibaryons \cite{PhysRevLett.123.162003}. We use the three set of $N_f = 2+1+1$ flavours HISQ gauge ensembles generated by MILC of lattice sizes $24^3\times 64$, $32^3\times 96$ and $48^3\times 144$ with lattice spacings 0.012, 0.0888 and 0.0582 fm, respectively \cite{Bazavov:2012xda}.
We use overlap action for the light to charm valence quarks and the valence propagators are calculated employing a gauge fixed wall source. As detailed in Refs. \cite{PhysRevD.99.031501,PhysRevLett.121.202002, Junnarkar:2018twb, PhysRevLett.123.162003} an NRQCD action, with improved coefficients is used for the bottom quarks. The strange quark mass is tuned by setting the unphysical pseudoscalar mass $\bar{s}s$ to 688 MeV, while the charm and bottom quark masses are set by equating the lattice determined spin-averaged kinetic masses of 1S quarkonia to their respective experimental values. 

\begingroup
\renewcommand*{\arraystretch}{1.5}
\begin{table}[h]
	\centering
	\begin{tabular}{cc|cc} \hline 
	   \multicolumn{2}{c}{Meson-meson type}    & \multicolumn{2}{c}{Antidiquark-diquark type}    \\
	  \hline 
	  1 & $(\bar{b}\gamma_5q_1)(\bar{c}\gamma_iq_2)$ & 4 & $([\bar{b}C\gamma_i\bar{c}][q_1C\gamma_5q_2])$  \\
	  \hline
	  2 & $(\bar{b}\gamma_iq_1)(\bar{c}\gamma_5q_2)$ & 5 & $([\bar{b}C\gamma_5\bar{c}][q_1C\gamma_iq_2])$\\
	  \hline
	  3 & $\epsilon_{ijk}(\bar{b}\gamma_jq_1)(\bar{c}\gamma_kq_2)$ & 6 & $\epsilon_{ijk}([\bar{b}C\gamma_j\bar{c}][q_1C\gamma_kq_2])$ \\
	  \hline
	\end{tabular}
	\caption{\label{tab:oper}The set of interpolating fields used for four-quark contents $\bar{b}\bar{c}q_1q_2$. [] refers to the color triplet (anti)diquark objects, whereas () refers to the color neutral objects. $C$ is the charge conjugation operator, and $\gamma$ refers to the Dirac matrices. The numbers on the left side of each operator are related to the operator ordering in GEVP. 
	Appropriate linear combinations of the above operators are taken for the projections to isospins I = 0 and I = 1/2.
	}
\end{table} 
\endgroup

The set of interpolating fields considered for this work with the valence quark  contents  $\bar{b}\bar{c}q_1q_2$, with $q_1,q_2  \in u,d,s$, are shown in Table~\ref{tab:oper}. These are all spatially-local operators. It is to be noted that along with the HQET-inspired operators, for which a heavy diquark has spin 1,   we also use the diquark-antidiquark type operators with the light diquark having a spin 1 keeping in mind that charm quark mass is much lighter than that of the bottom quark.

\section{Results}
We report our preliminary results on the energy spectra of $\bar{b}\bar{c}q_1q_2$ with various values of $m_{q_1}$ and $m_{q_2}$, while keeping the bottom and charm quark masses at their physical values. The relevant two-meson thresholds for the isoscalar axial-vector $\bar{b}\bar{c}ud$ channel are $B^*D, BD^*$ and $B^*D^*$, with $B^*D$ as the lowest one. For  $\bar{b}\bar{c}us$, corresponding two-meson thresholds are $B^*D_s, B_s^*D, BD_s^*, B_sD^*, B^*D_s^*$ and $B_s^*D^*$, where $B^*_sD$ is the lowest one. The matrices of two point correlation functions, $~C_{ij}(t)=\langle O_i(t+t_{\rm src})O_j^{\dagger} (t_{\rm src}) \rangle$, are constructed with the interpolators listed in Table \ref{tab:oper}, and are used to form a generalized eigenvalue problem $C_{ij}(t)v^{(n)}_j(t,t_0) = \lambda^{(n)}(t,t_0)C_{ij}(t_0)v^{(n)}_j(t,t_0)$. The energy spectra is then extracted from the large time behaviour of the eigenvalues $\lambda^{(n)}(t,t_0)$, with the exponential fits.

\begin{figure}[h]
  \begin{center}
\includegraphics[height=6.8cm,width=8.0cm]{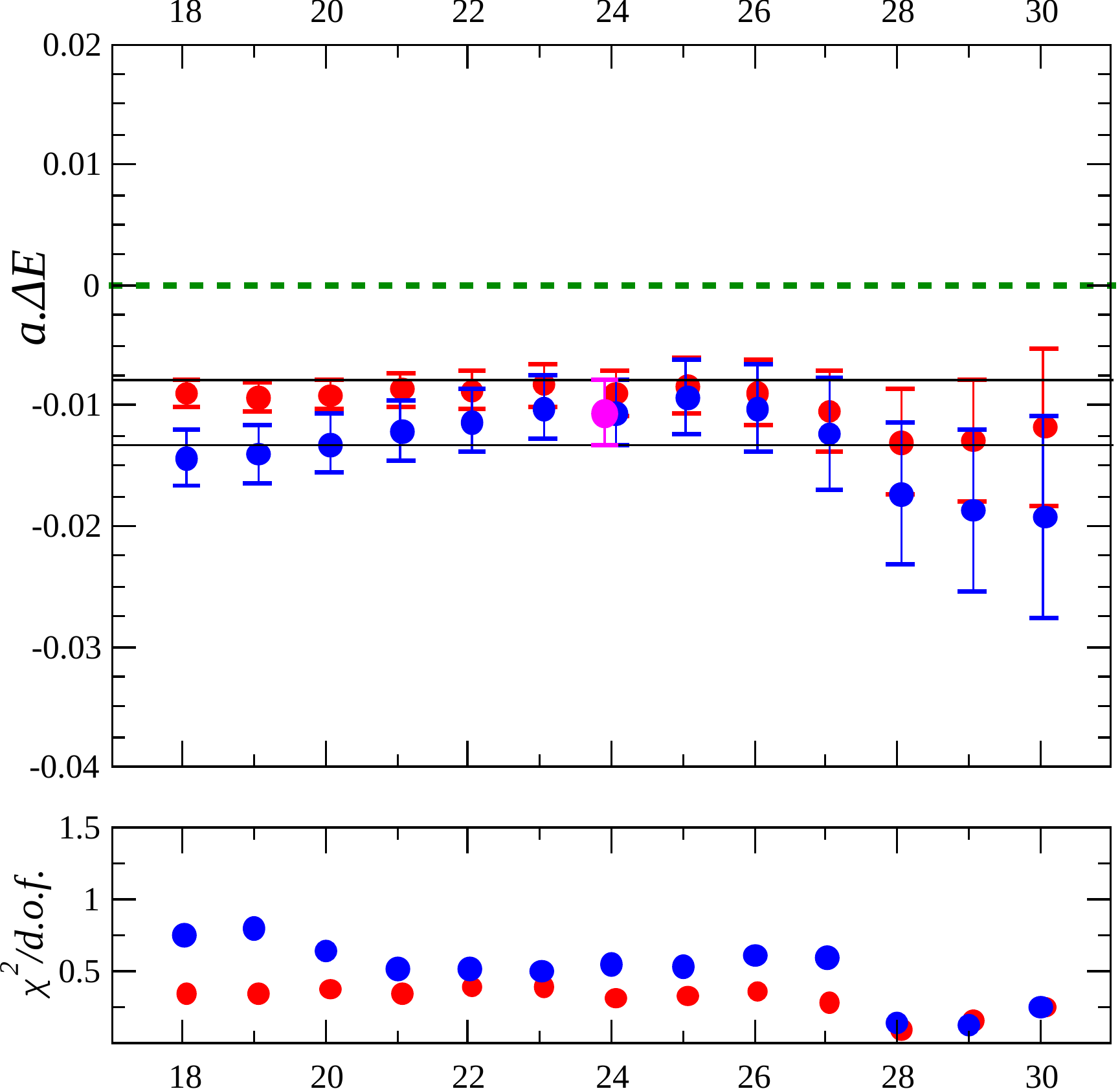}
\caption{Top: Energy splitting of the ground state with respect to the elastic threshold (shown by the dotted line) for the isoscalar axial-vector $\bar{b}\bar{c}us$ at the  quark mass $m_u$ corresponding to $m_{\pi}\sim 550$ MeV. The result is obtained on the fine lattice ensemble ($a = 0.0582$ fm) showing the dependence of the fitted result at times $t$. The red and blue circles are obtained by direct and ratio fitting methods. Bottom: $\chi^2$/d.o.f. for the respective fits.}\label{emass}
\end{center}
\end{figure}

We analyze various combination of operators to check the consistency in the ground state energy with the variation of the number of interpolating fields. In Fig.~\ref{emass} we show the energy splitting of the ground state from the elastic threshold for  $\bar{b}\bar{c}us$, at the quark mass $m_u$ corresponding to the pseudoscalar meson mass 550 MeV on the fine lattice ensemble ($a = 0.0582$ fm). The top part shows the energy difference with respect to the variation of fit-window and the bottom part is the corresponding $\chi^2$ per degrees of freedom for the corresponding fits.   

\begin{figure}
  \begin{center}
    \centering
    \includegraphics[height=7cm,width=10.0cm]{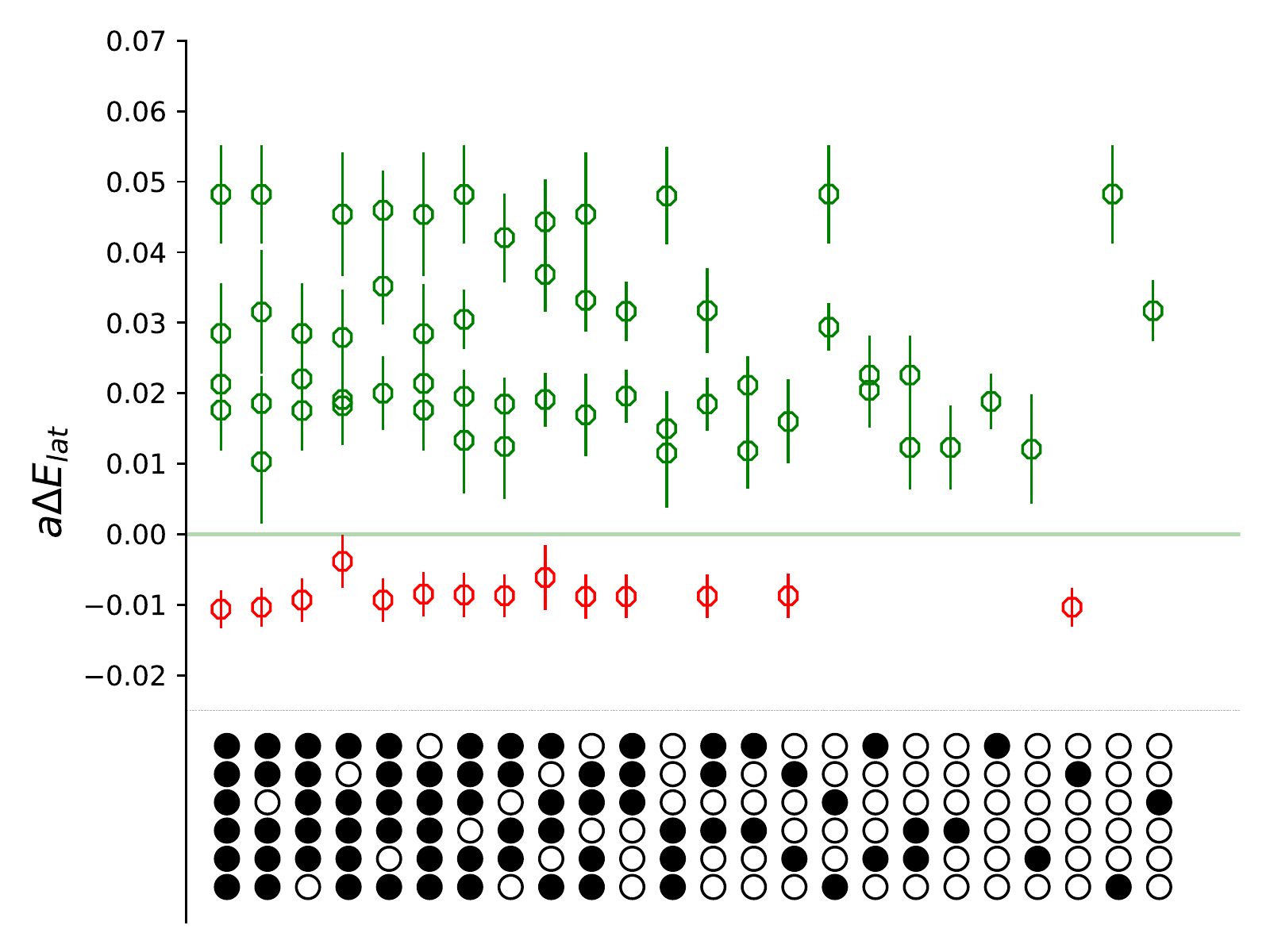}
\caption{A representative figure showing the basis dependence of the fitted energy differences from the elastic threshold (shown by the horizontal line set at zero). The circles below indicate whether an operator is included (filled) or not (empty) in a particular set of operators in the eigenvalue problem.}\label{op-basis}
\end{center}
\end{figure}

\begin{figure}
  \begin{center}
    \centering
    \includegraphics[height=6.8cm,width=10cm]{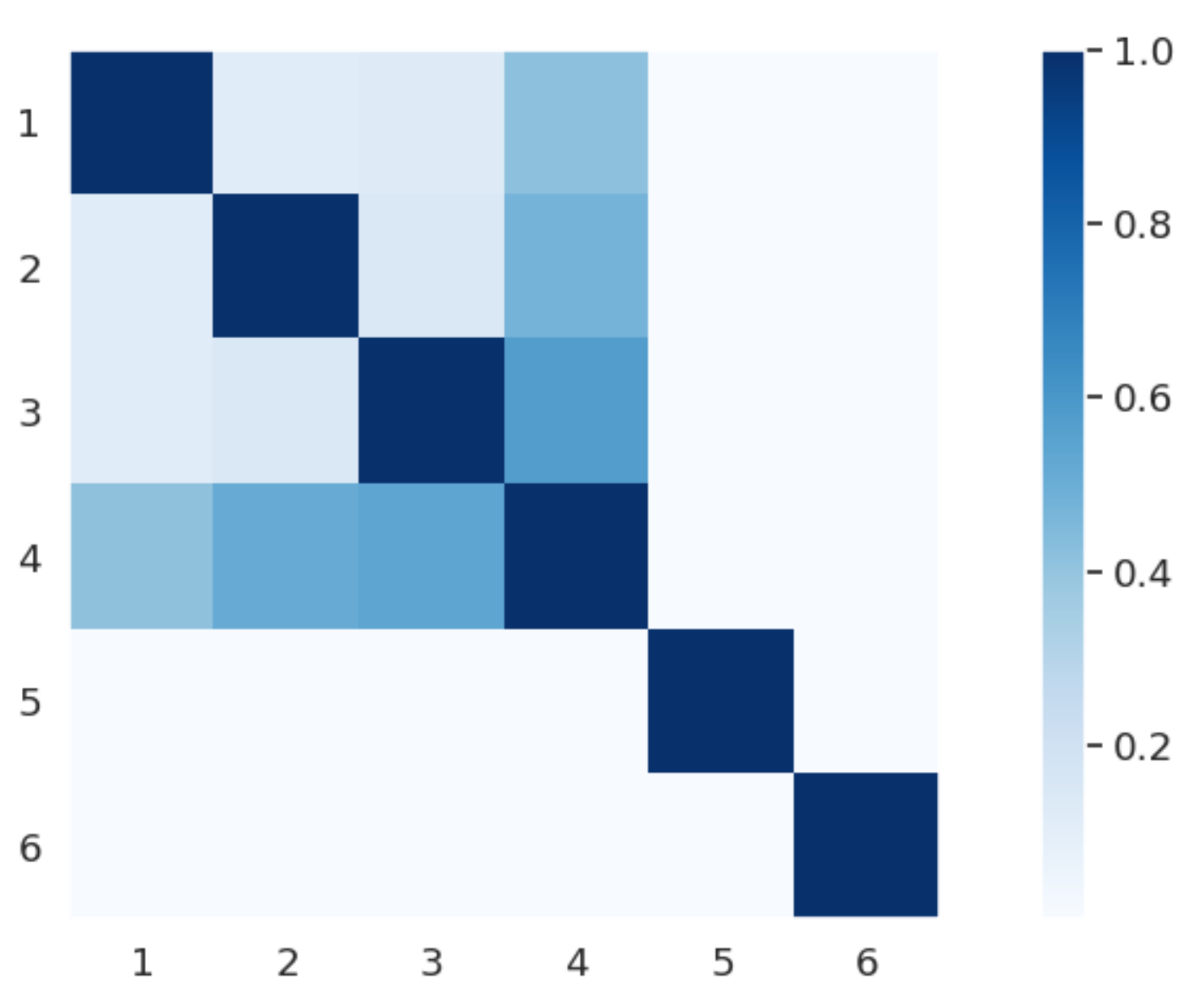}
\caption{The normalized correlation matrix, $C_{ij}/\sqrt{C_{ii}C_{jj}}$, at $t = 6$. This matrix plot, with relative color coding,  shows the relative strength of the correlation matrix elements with respect to the diagonal elements.
}\label{mat-plot}
\end{center}
\end{figure}

In Fig.~\ref{op-basis} the variation of the finite-volume spectrum, as extracted by the above procedure, is shown as a function of the basis of operators that we consider in the analysis. The filled and empty circles below represent whether an operator is included or not in the analysis. For example, in the farthest-left side the basis set has all filled circles, which signifies that we include all 6 operators in the sequence 1 to 6 as in Table~\ref{tab:oper}. The top part shows the energy difference as obtained in Fig.~\ref{emass} for a particular basis shown by column of circles shown vertically. It can be clearly seen that the inclusion of the operator 2 is essential to get an energy level below the lowest threshold. This particular figure is for the case of $\bar{b}\bar{c}us$ on the fine lattice ensemble ($a = 0.0582$ fm) at the physical bottom and charm quark mass and at $m_u$ corresponding to the pseudoscalar meson mass of 550 MeV. To see the relative strength of the correlation matrix elements with respect to the diagonal elements
 one can define a normalized correlation matrix, $C_{ij}/\sqrt{C_{ii}C_{jj}}$.
 In Fig.~\ref{mat-plot} we show a matrix plot representing that cross-correlation between various operators. It is interesting to see that operator 4, the HQET-inspired operator for which a heavy diquark has spin 1, is the only antidiquark-diquark type operator that has strong cross-correlations with meson-meson operators. 

\begin{figure}
  \begin{center}
    \includegraphics[height=6.5cm,width=7.5cm]{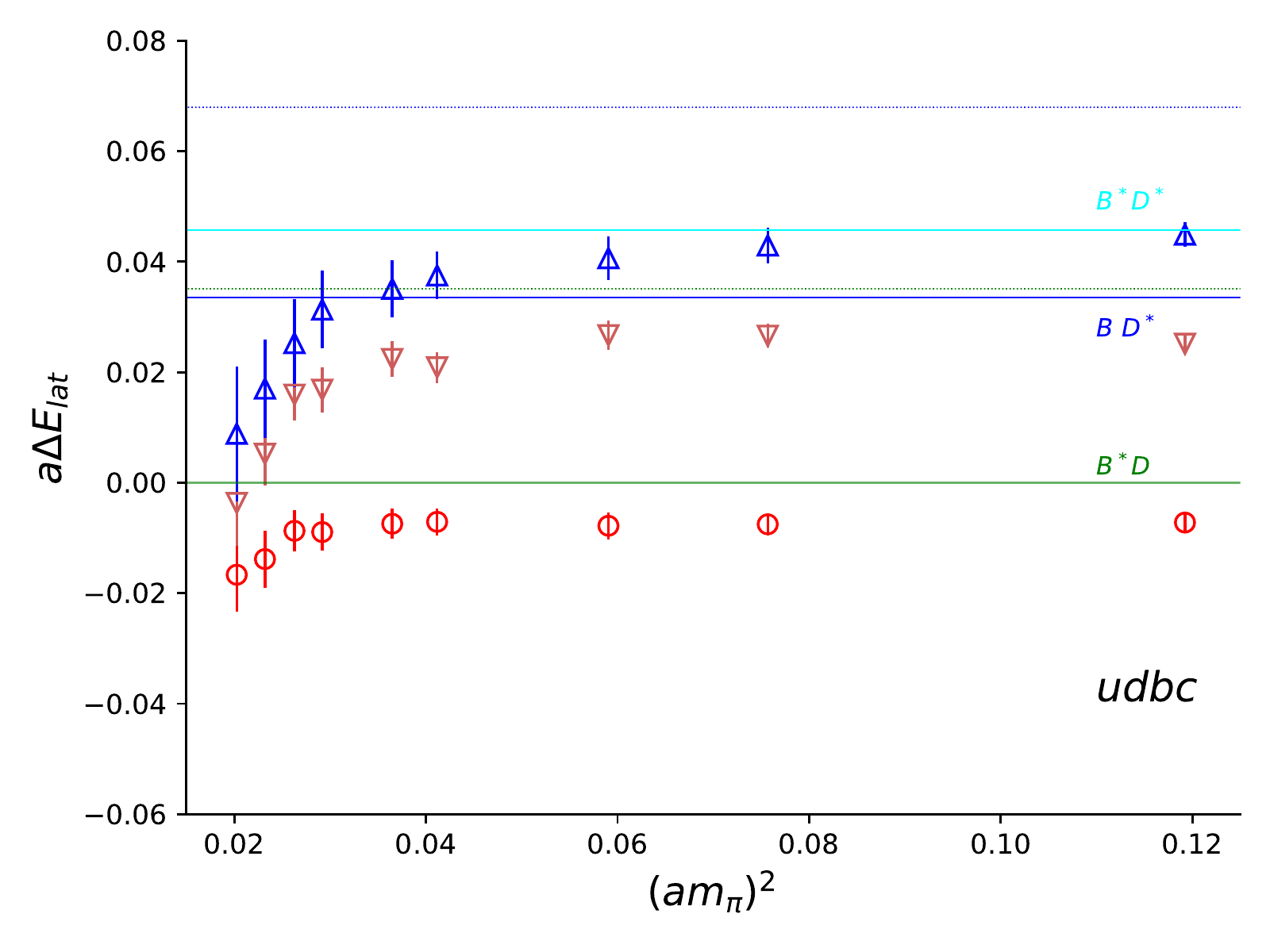}
     \includegraphics[height=6.5cm,width=7.5cm]{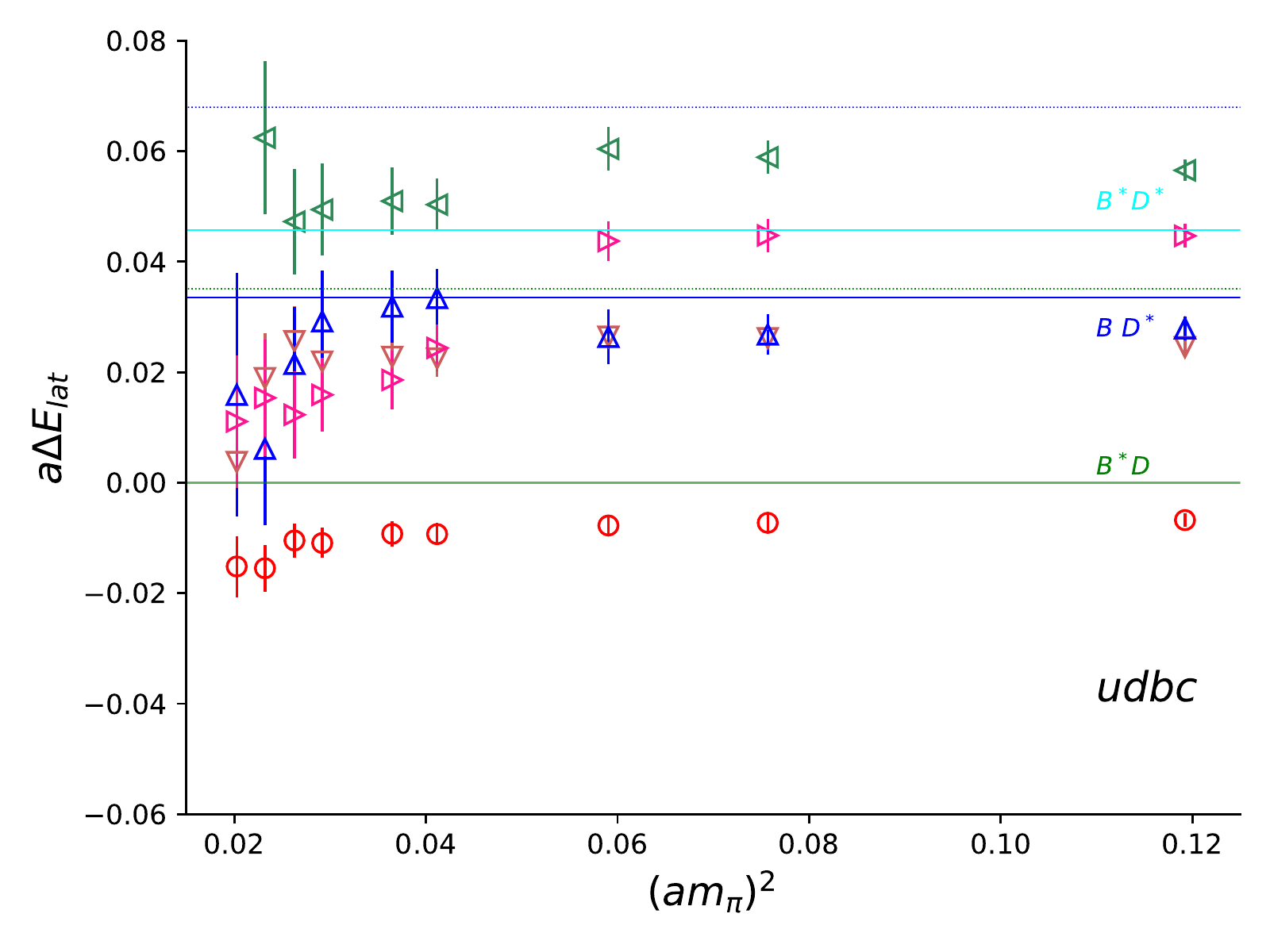}    
\caption{The finite-volume energy spectra of the isoscalar axial-vector $\bar b\bar cud$ in terms of their splitting from the elastic threshold, at various values of $m_u = m_d$ corresponding to pion masses shown in the $x$-axis. The horizontal lines refer various non-interacting levels with the elastic threshold set at zero.  Left: For the basis including operators 1, 2, and 4. Right: Same as on the left, but for the full operator basis. }\label{udbc}
\end{center}
  \end{figure}

We perform similar analysis for $\bar{b}\bar{c}q_1q_2$ at a number of light quark masses ($m_{q_1} = m_{q_2} = m_q$) while keeping the charm and the bottom quark masses at their physical values. In Fig.~\ref{udbc} we show the results for two set of operator basis -- when only operators 1, 2 and 4 are included (left), and when all 6 operators are included (right). The vertical-axis represents the energy difference of the extracted energy levels from the elastic $B^*D$-threshold while the horizontal-axis is the pion mass corresponding to the different light-quark masses ($m_{q_1} = m_{q_2} = m_q$). These results are obtained at the finest ensemble ($a = 0.0582$ fm) and show the presence of an energy level below the  lowest $B^*D$-threshold for all values of the light quark masses even above the strange quark mass. This energy difference ($\Delta E$) increases as $m_{q_1}, m_{q_2}$ approach their physical values. On the higher side one needs to investigate whether $\Delta E$ reduces further and approach its elastic threshold limit at  $m_{q_1} = m_{q_2} = m_c$ or $m_b$. A preliminary estimate for $|\Delta E|_{\bar{b}\bar{c}ud}$ is found to be $\sim  20-40$ MeV at the physical quark masses.

\begin{figure}
  \begin{center}
    \includegraphics[height=6.5cm,width=7.5cm]{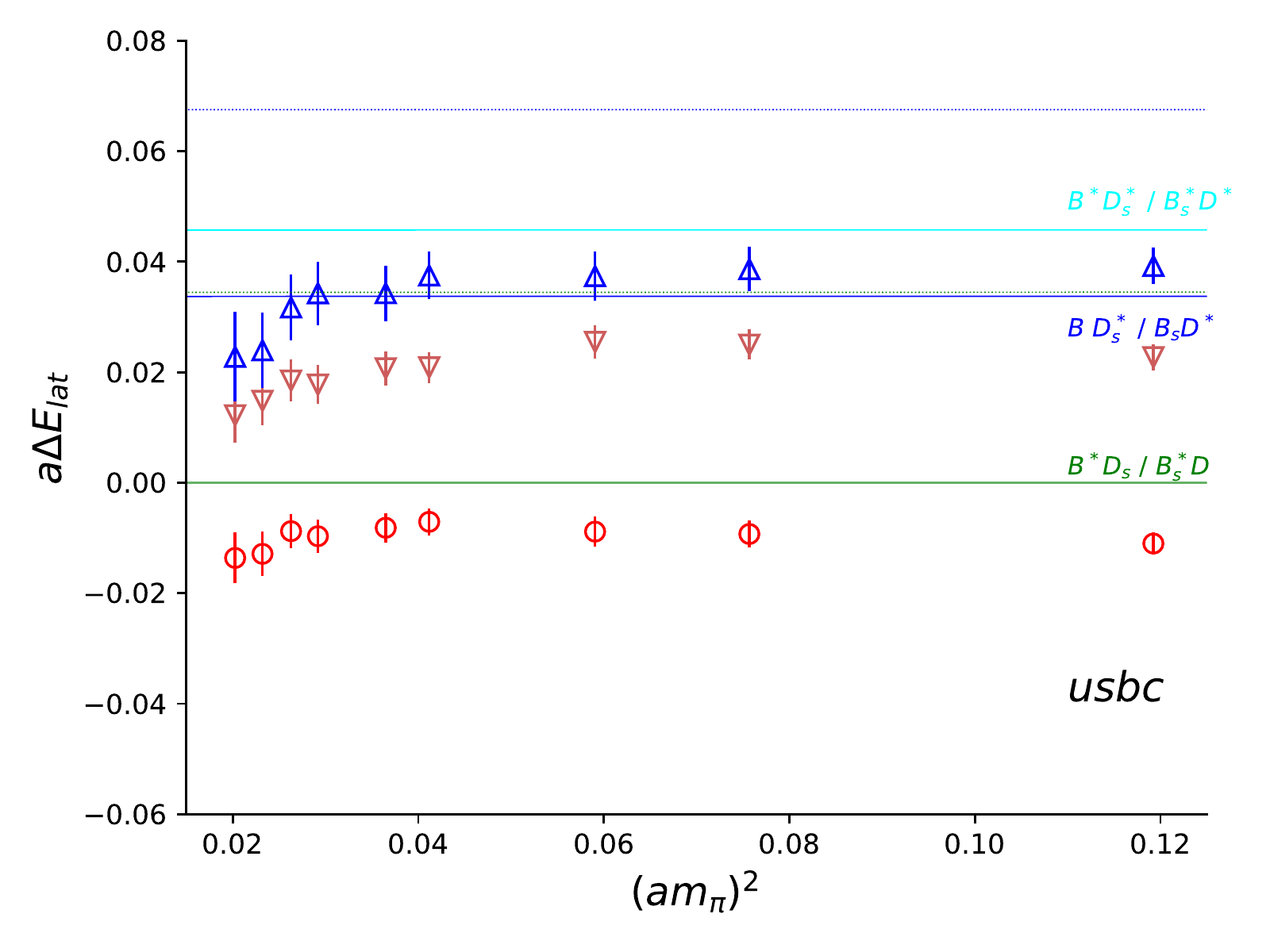}
     \includegraphics[height=6.5cm,width=7.5cm]{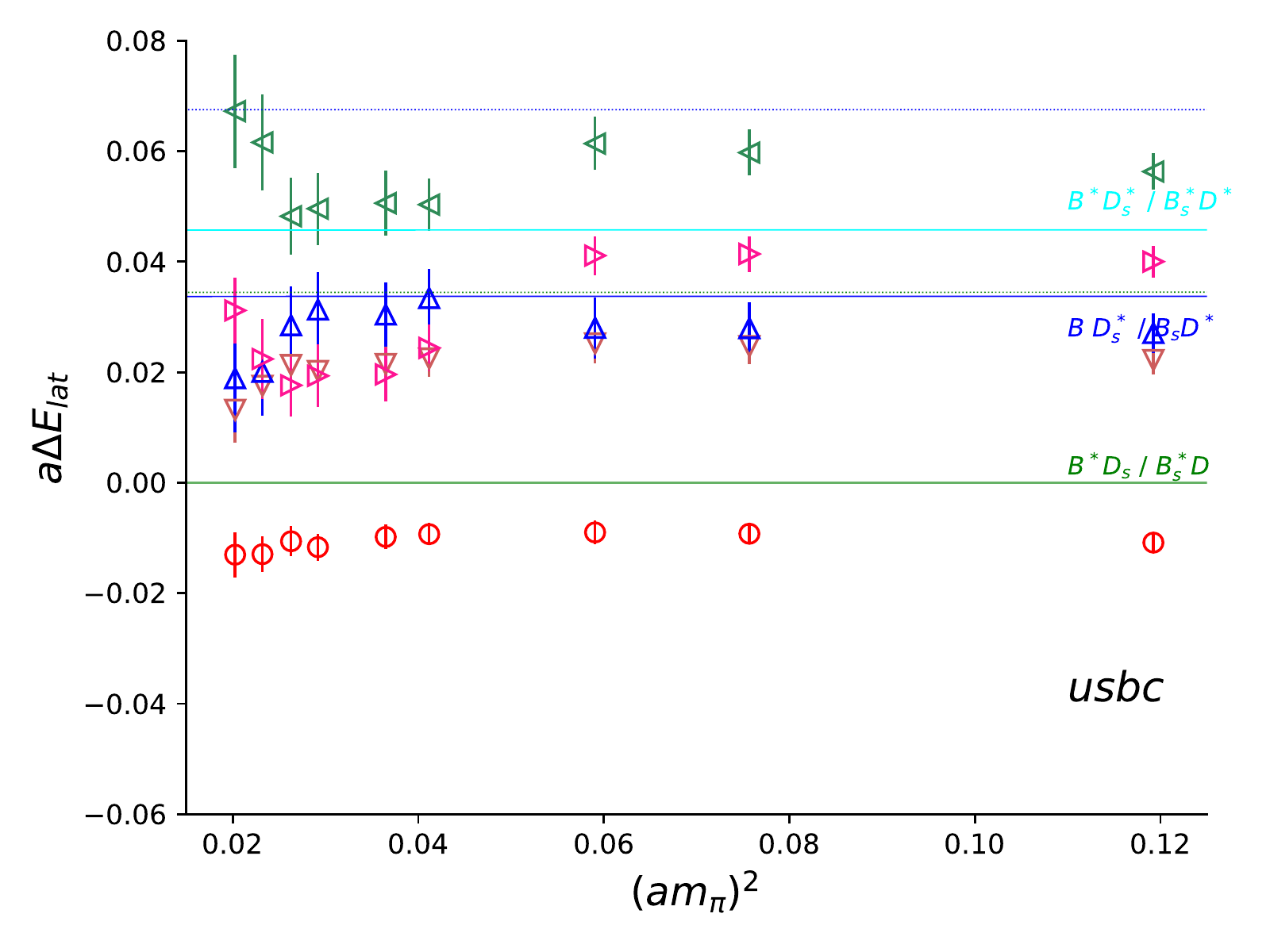}    
\caption{Same as in Figure \ref{udbc}, but for the isoscalar axial-vector $\bar b\bar cus$ channel}\label{usbc}
\end{center}
  \end{figure}

In Fig.~\ref{usbc} we show the similar results for the four-quark configurations   $\bar{b}\bar{c}qs$ where the strange, charm and bottom quark masses are kept at their physical values while the light quark mass $m_{q}$ (equivalently the pion mass) is varied. Here also we see an energy level below the lowest threshold $B^*_sD$, that is, the extracted $\Delta E$ is non-zero and negative within the statistics. Here again,  as in the case of $\bar{b}\bar{c}q_1q_2$, we see the similar pion mass dependence of $\Delta E$, that is, smaller the light quark mass larger is the value of $|\Delta E|$.

\begin{figure}[h]
  \begin{center}
\includegraphics[height=6.8cm,width=9.0cm]{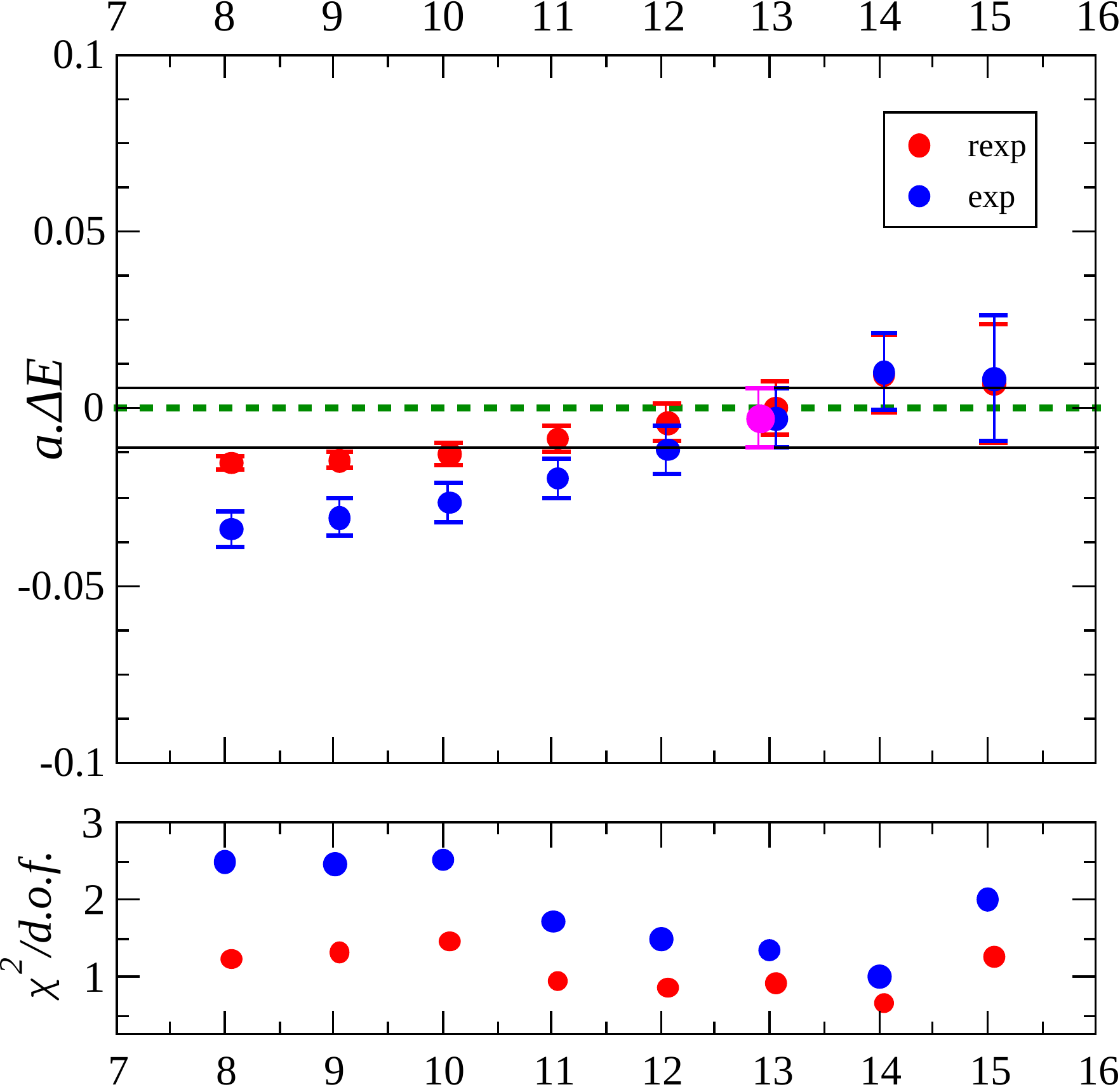}
\caption{Same as Fig. \ref{emass} for $\bar{b}\bar{c}us$ four-quarks, but at a coarser lattice spacing ($a = 0.12$ fm) and at a light quark mass ($m_u$) corresponding to the pion mass 340 MeV. Compared to the fine lattice results the energy splitting shown above is smaller and is consistent with zero.}\label{emass_24q}
\end{center}
\end{figure}

Next, to see the lattice spacing dependence of the extracted energy differences ($\Delta E$) we perform similar study on our coarsest lattice ensemble ($a = 0.12$ fm). In Fig. \ref{emass_24q}
we show a representative plot to illustrate our results for $\bar{b}\bar{c}us$ at a light quark mass ($m_u$) corresponding to the pion mass 340 MeV.
This plot shows that the energy splitting of the lowest energy level from the elastic threshold, 
similar to the one presented in Fig.\ref{emass}, on a coarser lattice is smaller and consistent with zero within the statistics. We plan to further study the lattice spacing dependence of the energy spectra on a larger number of configurations and also with another lattice spacing at $a = 0.09$ fm. 

\section{Conclusions and outlook}
Motivated by the discovery potential of four-quark bound states with the valence quark configurations $\bar{b}\bar{c}ud/\bar{b}\bar{c}us$,
we have performed a lattice QCD calculation to study the energy spectra of these systems. We report preliminary results for the cases with $J = 1$ and isospins $I = 0$ and $1/2$.  Our preliminary results indicate the presence of an energy level below the respective elastic threshold both for $\bar{b}\bar{c}ud$ and $\bar{b}\bar{c}us$. We also find that the energy splitting between the ground state from the elastic threshold is dependent on the lattice spacing and it increases as one approaches the continuum limit. This could be related to the cut-off effects of the energy spectra. Hence one needs to use a finer lattice so that $m_ca$ value for the charm quark is small. Our preliminary estimate, at the fine lattice ensemble, of this energy difference for $\bar{b}\bar{c}ud$ is in the range of 20-40 MeV below its elastic threshold.  A detailed investigation on the volume dependence of these finite-volume energy levels is further required to make concrete inferences on the existence of bound states in these channels. We plan to perform such a study in future.


\begin{acknowledgments}
This work is supported by the Department of Atomic Energy, Government of India, under Project Identification Number RTI 4002.
We are thankful to the MILC collaboration and in particular to S. Gottlieb for providing us with the HISQ lattices. Computations are carried out on the Cray-XC30 of ILGTI, TIFR,   and on the Pride/Flock/Zeal clusters of the Department of Theoretical Physics,
TIFR. 
\end{acknowledgments}

\bibliographystyle{JHEP}
\bibliography{lat21}



\end{document}